\documentclass[preprint,nofootinbib,aps,superscriptaddress]{revtex4}

\usepackage{graphicx}
\usepackage{amsmath}

\newcommand{\beq}{\begin{equation}}
\newcommand{\eeq}{\end{equation}}
\newcommand{\beqa}{\begin{eqnarray}}
\newcommand{\eeqa}{\end{eqnarray}}
\newcommand{\nn}{\nonumber}
\newcommand{\PSbox}[3]{\mbox{\rule{0in}{#3}
\includegraphics{#1}\hspace{#2}}}

\def\OMIT#1{{}}

\begin{document}

\preprint{\vbox{ 
\hbox{CALT-68-2517} 
\hbox{hep-ph/0408263}}}

\vspace*{1.25cm}

\title{\boldmath Negative-Parity Heavy Pentaquark States in $1/N_c$\vspace{0.2cm}}

\author{Margaret E.\ Wessling\,}\email{wessling@theory.caltech.edu}
\affiliation{California Institute of Technology, Pasadena, CA 91125
        \\}

%\pacs{12.10.Dm, 12.10.Kt, 98.80.Cq}

\begin{abstract}
  
The $1/N_c$ expansion for negative-parity heavy pentaquarks is developed using the formalism introduced for excited baryons in large $N_c$.  Relations are found between the mass splittings of these pentaquarks and those of nonexotic baryons.

\end{abstract}
\maketitle

Experimental evidence for the $\Theta^+(1540)$ pentaquark \cite{nakano} has occasioned much recent theoretical interest in exotic baryons.  There has also been a report, from the H1 Collaboration, of a heavy anticharmed analogue, the  $\Theta_c$ \cite{aktas}.  The existence of these five-quark states has not been firmly established; ZEUS did not find the heavy pentaquark seen by H1 \cite{lipka}, and a number of experiments have searched for the $\Theta$ with null results \cite{aubert,hicks}.  Whatever the experimental consensus on the $\Theta$ and $\Theta_c$ turns out to be, the existence of pentaquarks remains an intriguing possibility; nothing in QCD appears to rule them out. 

The $1/N_c$ expansion has recently been extended to exotic baryons, including partners of the $\Theta$ \cite{cohen, manohar} and heavy pentaquarks in which the antiquark is a $\bar c$ or a $\bar b$  \cite{manohar}.  This work assumed that the pentaquark states are in the completely symmetric representation of spin-flavor $SU(6)$ and thus have positive parity.  Such an assumption is sensible because it has been shown, in the context of a constituent quark model \cite{glozman}, that the hyperfine flavor-spin interactions between quarks in a hadron are most attractive for completely symmetric states.  However, in order to satisfy Fermi statistics, a positive-parity pentaquark would need to have one quark in an orbitally excited $\ell = 1$ state; it is not clear whether the resulting P-wave energy would always be sufficiently offset by the attractive flavor-spin interactions to make the positive-parity pentaquarks lighter than their negative-parity counterparts.  Inspired by the diquark model of Jaffe and Wilczek \cite {Jaffe:2003sg}, we have previously considered heavy pentaquarks in a mixed-symmetry representation of $SU(6)$, which can be in an S-wave state, and argued that these may be stable against strong decays \cite{stewart}.  Here we consider such states in the context of a $1/N_c$ expansion.

In the $N_c \rightarrow \infty$ limit, baryons form irreducible representations of contracted spin-flavor $SU(6)_c$; for finite $N_c$, this symmetry is broken, generating mass splittings within each representation.  The symmetry breaking can be parameterized using polynomials in the $SU(6)$ generators:
%%%%%%%%%%%%%%%%%%%%%%%%%%%%%%%%%%%%%%%%%
\begin{align} \label{generators}
  S^i & \equiv  q^\dag (\frac{\sigma^i} {2} \otimes \openone)q \nn \\
  T^a & \equiv  q^\dag (\openone \otimes \frac{\lambda^a}{2})q \nn \\
  G^{ia} & \equiv  q^\dag  (\frac{\sigma^i}{2} \otimes \frac{\lambda^a}{2})q
\end{align}
%%%%%%%%%%%%%%%%%%%%%%%%%%%%%%%%%%%%%%%%%
where $q^\dag$ and $q$ are quark creation and annihilation operators, $\sigma^i$ are the Pauli matrices, and $\lambda^a$ are the Gell-Mann matrices.  An $n$-body operator, which acts on $n$ quark lines in a baryon, comes with a factor $N_c^{1-n}$.  The generator $G^{ia}$ sums coherently over all the quark lines and hence is order $N_c$.  $T^a$ may also sum coherently when three or more flavors are considered.  (When the discussion is limited to two flavors, as in \cite{carlson99}, the isospin is fixed in the large $N_c$ limit, so $T^a$ is order 1.) Thus a given $n$-body operator contributes at order $N_c^{1-n-m-p}$, where $m$ is the number of times $G^{ia}$ appears and $p$ is the number of times $T^a$ appears.  To describe mass splittings, one constructs all possible scalar operators up to a given order in $1/N_c$; each such operator appears in the expansion with an unknown coefficient of order unity.  Depending on the symmetry of the baryon states under consideration, there may be operator reduction rules allowing some operators to be eliminated; the rules for completely symmetric states are given in \cite{dashen}.  (See \cite{march} for a different approach to $1/N_c$ calculations.)

\begin{figure}[h]
\PSbox{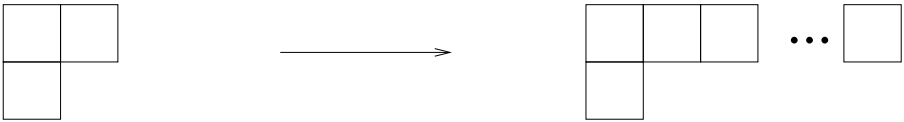 hoffset=40 voffset=10 hscale=80
vscale=80}{6.8in}{2.2in}
\caption{Extension of the excited baryons to large $N_c$.  The top row of the tableau on the right has $N_c - 1$ boxes.}
\label{fig:c1}
\end{figure}

The $1/N_c$ expansion for excited baryons provides a model for working with states of mixed spin-flavor symmetry.  References \cite {carlson99,goity,pirjol} study excited baryons in the {\bf 70} of $SU(6)$; they generalize this representation for $N_c > 3$ as shown in the Young diagram in figure \ref{fig:c1}. 

%%%%%%%%%%%%%%%%%%%%%%%%%%%%%%%%%%%%
In this picture, a baryon contains one excited quark, with angular momentum $\ell = 1$, and $N_c - 1$ ``core'' quarks, which are completely symmetric in spin-flavor $SU(6)$.  The expansion is made using two sets of $SU(6)$ generators: $s^i$, $t^a$, $g^{ia}$, acting on the excited quark; and $S^i_c$, $T^a_c$, $G^{ia}_c$, acting on the core.  The reduction rules for these operators are determined in \cite {carlson99}.
 
\begin{figure}[h]
\PSbox{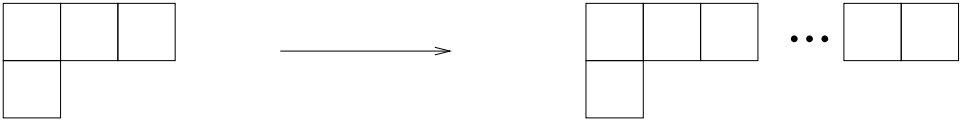 hoffset=40 voffset=10 hscale=80
vscale=80}{6.8in}{2.2in}
\caption{Extension of the negative-parity pentaquarks to large $N_c$.  The top row of the tableau on the right now has $N_c$ boxes.}
\label{fig:c2}
\end{figure}

We wish to examine exotic negative-parity baryons containing $N_c + 1$ light quarks and one heavy antiquark, which can be extended to large $N_c$ in a similar manner, as shown in figure \ref{fig:c2}.  
%%%%%%%%%%%%%%%%%%%%%%%%%%%%%%
%%%%%%%%%%%%%%%%%%%%%%%%%%%%%%%
Here too, it makes sense to construct operators from two different sets of generators.  The term ``excited'' does not apply in this case, because the pentaquarks have no orbital angular momentum.  However, dividing the states into $N_c$ symmetrized quarks (which we will continue to call the ``core''), plus one extra, still captures their symmetry properties in a useful way.   The same operators and reduction rules constructed for the excited baryons in \cite{carlson99,goity,pirjol} may be used to describe the negative-parity pentaquarks.  In fact, the situation simplifies significantly in the pentaquark case, because the seven operators depending on $\ell$ all vanish.  We are left, at order $1/N_c$, with six linearly independent operators:\footnote{The expansion should also contain operators depending on the exoticness $E$, such as $E \openone$ and $\frac{1}{N_c} E^2$.  However, since $E=1$ for all the pentaquark states, these operators do not tell us anything new about the mass splittings.}

\begin{align}
O_1 & \equiv  N_c \openone, \nn \\ 
O_2 & \equiv  \frac{1}{N_c} S_c^2, \nn \\ 
O_3 & \equiv  \frac{1}{N_c} s^iS_c^i, \nn \\  
O_4 & \equiv  \frac{1}{N_c} t^aT_c^a, \nn \\
O_5 & \equiv  \frac{2}{N_c^2} t^a \{ S_c^i,G_c^{ia} \}, \nn \\
O_6 & \equiv  \frac{1}{N_c^2}g^{ia}S_c^iT_c^a.
\end{align}

The {\bf 210} of $SU(6)$, which describes the $N_c + 1$ light quarks,  can be decomposed into flavor $\otimes$ spin to give seven different multiplets\footnote{The superscript $P$ indicates pentaquark states. Later we will use $E$ to denote excited baryons and $N$ for normal baryons.} (see, e.g., \cite{itzykson}): ${\bf 15}^{\prime P} _1$, ${\bf 15}_2^P$,  ${\bf 15}_1^P$, ${\bf 15}_0^P$,  ${\bf \bar 6}_1^P$,  ${\bf 3}_1^P$, and  ${\bf 3}_0^P$.  In extending this decomposition to large $N_c$, the spin of each state remains fixed.  The flavor representations change with $N_c$; however, we will always use the $N_c = 3$ values for notational purposes in this paper. (We will also continue to use the term ``pentaquark'' for the large-$N_c$ analogues of such states.)  Appendix B contains diagrams of the four different flavor representations.

The ${\bf 3}_0^P$ is the multiplet called $T_a$ or $R_a$ in \cite{stewart}; the three states have the flavor content $\bar Q uuds$, $\bar Q udds$, and $\bar Q udss$.  Interestingly, the large-$N_c$ version of this state can still be thought of in terms of the diquark model.  In this view, a ``diquark'' for arbitrary $N_c$ still consists of two quarks combined antisymmetrically in color and flavor; it may be written $\phi^a_{[\alpha \beta]}$, where $a$ is a flavor index and $\alpha$, $\beta$ are antisymmetrized color indices.  The full state is then
%%%%%%%%%%%%%%%%%%%%%%%%%%%%%%%%
\beq
T_a^{d_1...d_{(N_c-3)/2}} = \delta^\alpha_\beta \epsilon_{abc} \epsilon^{\gamma_1\gamma_2...\gamma_{N_c}} \bar Q^\alpha \phi^b_{\beta \gamma_1} \phi^c_{\gamma_2 \gamma_3} \phi^{d_1}_{\gamma_4 \gamma_5}...\phi^{d_{(N_c-3)/2}}_{\gamma_{N_c-1} \gamma_{N_c}}
\eeq
%%%%%%%%%%%%%%%%%%%%%%%%%%%%%%%
The ${\bf 3}_0^P$ is the only one of the seven multiplets that can be constructed using Jaffe and Wilczek's original spin 0, flavor ${\bf \bar 3}$ diquarks.  However, if we also allow spin 1, flavor ${\bf 6}$ diquarks--called tensor diquarks  in \cite{shuryak} and ``bad'' diquarks in \cite{jaffexotic}--many other multiplets become possible.  The remainder of this paper makes no reference to the diquark model; the results are model-independent.\footnote{There is a rather subtle issue regarding model-independence: do the results of this analysis follow from large-$N_c$ QCD alone, or do they depend on the large-$N_c$ quark model?  The ordinary ground-state baryons are stable in the large-$N_c$ limit, with properties entirely determined by symmetry; the quark model in this case is just a convenient way of counting states, introducing no dynamical assumptions beyond large-$N_c$ QCD.  However, this is not true for the excited baryons; they have widths that go as $N_c^0$, and so using the quark model for them does introduce new dynamical assumptions.  References \cite{cohen2} address this problem in detail. There are reasons to suspect that the exotic baryon states considered here are in fact stable at large $N_c$, in which case the present results would be truly model-independent. The rigorous proof of this will be left to a future publication.}

It is straightforward to determine what the ``core'' of each state should look like: $S_c = S_{total} \pm \frac{1}{2}$, and the core flavor representation can be written in Dynkin index notation as $(\lambda, \mu) = (2S_c, \frac{N_c-2S_c}{2})$.  For six of the seven multiplets, there is only one possible value of $S_c$.  In particular: $S_c = \frac{1}{2}$ for the ${\bf 15}_0^P$,  ${\bf \bar 6}_1^P$,  ${\bf 3}_1^P$, and  ${\bf 3}_0^P$ states;  $S_c = \frac{3}{2}$ for the ${\bf 15}^{\prime P} _1$ and ${\bf 15}_2^P$ states.  The ${\bf 15}_1^P$ state is somewhat more complicated, because the flavor-spin decomposition of the totally symmetric representation of $SU(6)$ also contains a ${\bf 15}_1$.  The correct core is a linear combination of $S_c = \frac{3}{2}$ and  $S_c = \frac{1}{2}$, whose coefficients can be determined using Casimir operators.  Reference \cite{carlson99} finds the analogous coefficients for total spin $S$ and a core of $N_c - 1$ quarks; we may simply use their result with $S = 1$ and $N_c \rightarrow N_c + 1$: 

\beq
  \sqrt{\frac{N_c+5}{3(N_c+1)}} \left | S_c = \frac{3}{2} \right > - \sqrt{\frac{2(N_c-1)}{3(N_c+1)}} \left | S_c = \frac{1}{2} \right >
\eeq

Evaluating the matrix elements of some of these operators, particularly $O_5$, is rather a nontrivial task; it cannot be done by a simple $N_c \rightarrow N_c + 1$ substitution.  One method of evaluation is to construct the wavefunction for each state as in section II of \cite {goity}, and use the Wigner-Eckart theorem to express each matrix element in terms of Clebsch-Gordan coefficients and the reduced matrix elements of $S_c^i$, $T_c^a$, $G_c^{ia}$, $s^i$, $t^a$, and $g^{ia}$.  The $SU(2)$ Clebsch-Gordan coefficients may be calculated in, e.g., Mathematica, or looked up in any of many published tables; analytic formulas for the necessary $SU(3)$ coefficients appear in \cite{hecht,vergados}.  One may also use the bosonic operator method described in \cite{manohar04}.  Explicit values for the relevant matrix elements appear in Appendix A.

We arrive at the following mass relation among the negative-parity heavy pentaquarks:

\beq \label{mass1}
  {\bf 15}_2^P - {\bf 15}^{\prime P} _1 = 2({\bf \bar 6}_1^P - {\bf 15}_0^P) + \mathcal{O}(1/N_c^3)
\eeq
Note that this relation holds to order $1/N_c^2$. In addition to $O_5$ and $O_6$, the operator $O_{11} \equiv \frac{1}{N_c^3}S_c^2t^aT_c^a$ also contributes at $\mathcal{O}(1/N_c^2)$, and the relation remains true when this contribution is added.

The pentaquark states can also be related to the excited baryons, using the results from \cite{goity}, and to the ground state octet and decuplet baryons.  (Note that the core of the octet is a linear combination of $S_c =1$ and $S_c = 0$,with coefficients given in \cite{carlson99}.)  The mass relations include 
%%%%%%%%%%%%%%%%%%%%%%%%%%%%%%%%%%%%%%%%%
\begin{align}
{\bf \bar 6}_1^P - {\bf 15}_0^P  =   \frac{2}{3}(\left < ^4{\bf 8}^E \right > - \left < ^2{\bf 10}^E \right >) + \mathcal{O}(1/N_c^2) \label{split1} \\
({\bf 3}_1^P-{\bf 3}_0^P)-\frac{7}{11}({\bf 15}_2^P-{\bf 15}_1^P)+\frac{17}{11}({\bf 15}_1^P- {\bf 15}_0^P) &  = \nn \\
\frac{2}{11}( \left < ^2{\bf 10}^E \right >-\left < ^2{\bf 8}^E \right >) 
& +\frac{4}{11}( \left < ^4{\bf 8}^E \right >-\left < ^2{\bf 8}^E \right >)+ \mathcal{O}(1/N_c^2)  \label{split2} \\
\frac{11}{4}({\bf 15}^{\prime P}_1-{\bf 15}_2^P)+\frac{28}{11}({\bf 15}_2^P-{\bf 15}_1^P)-\frac{79}{11}({\bf 15}_1^P-{\bf \bar 6}_1^P)-2({\bf 15}_1^P-{\bf 3}_0^P) & = \nn \\ 
({\bf 10}_{3/2}^N-{\bf 8}_{1/2}^N) 
&  +2\left < ^2{\bf 1}^E \right >+\frac{13}{11}\left < ^2{\bf 8}^E \right > \nn \\
-\frac{35}{11}\left < ^2{\bf 10}^E \right > & + \mathcal{O}(1/N_c^2) \label{split3}
\end{align}
%%%%%%%%%%%%%%%%%%%%%%%%%%%%
Here ${\bf 10}_{3/2}^N$ and ${\bf 8}_{1/2}^N$ are the nonexotic octet and decuplet;  $\left < ^2{\bf 10}^E \right >$, $\left < ^4{\bf 8}^E \right >$, $\left < ^2{\bf 8}^E \right >$, and $\left < ^2{\bf 1}^E \right >$ are spin averages of the excited baryons:\footnote{There is some ambiguity involved in identifying the ${\bf 8}_{1/2}^E$ and ${\bf 8}_{3/2}^E$ multiplets with physical states, because the values of the mixing angles are not known.  The results above are calculated assuming zero mixing.}
%%%%%%%%%%%%%%%%%%%%%%%%%%%%%%
\begin{align}
  \left < ^2{\bf 10}^E \right > & =  \frac{1}{3}(^2{\bf 10}_{1/2}^E+2(^2{\bf 10}_{3/2}^E)) \nn \\
   \left < ^2{\bf 8}^E \right > & =  \frac{1}{3}(^2{\bf 8}_{1/2}^E+2(^2{\bf 8}_{3/2}^E)) \nn \\
    \left < ^2{\bf 1}^E \right > & =  \frac{1}{3}(^2{\bf 1}_{1/2}^E+2(^2{\bf 1}_{3/2}^E)) \nn \\
     \left < ^4{\bf 8}^E \right > & =  \frac{1}{6}(^4{\bf 8}_{1/2}^E+2(^4{\bf 8}_{3/2}^E)+3(^4{\bf 8}_{5/2}^E))
\end{align}
%%%%%%%%%%%%%%%%%%%%%%%%%%%%

The order $N_c$ contribution to the pentaquark mass is about 1 GeV, so we estimate $\mathcal{O}(1/N_c^2)$ corrections to be of order 30 MeV, and $\mathcal{O}(1/N_c^3)$ to be of order 10 MeV.  Using the Particle Data Group values \cite{pdg} for the masses of the nonexotic baryons, we can give numerical estimates for the right-hand sides of equations (\ref{split1}) - (\ref{split3}).  The mass difference $\left < ^4{\bf 8}^E \right > - \left < ^2{\bf 10}^E \right >$ is quite small, about 4 MeV; the error in eq. (\ref{split1}) is estimated to be considerably larger than this.    Thus eq. (\ref{split1}) indicates that the ${\bf \bar 6}_1^P$ and ${\bf 15}_0^P$ masses are close together, with a splitting of 4 $\pm$ 30 MeV, but cannot tell us which one is heavier.  The same applies to the ${\bf 15}_1^{\prime P}$ and ${\bf 15}_2^P$ masses, by eq. (\ref{mass1}).  The right-hand side of eq. (\ref{split2}) is about 83 MeV, again with an error of $\pm$ 30 MeV.  If all three pentaquark splittings on the left-hand side were equal, each would be about 40 $\pm$ 30 MeV.  Based on the estimate in \cite{stewart} that the isospin $\frac{1}{2}$ members of the ${\bf 3}_0^P$ should have mass 2580 MeV, this very rough guess suggests that the corresponding members of the ${\bf 3}_1^P$ would have mass 2620 $\pm$ 30 MeV, meaning that they would also be too light to decay to a $D_s$ plus a proton.    The right-hand side of eq. (\ref{split3}) comes to -260 $\pm$ 30 MeV. 

The light quarks combine with the heavy antiquark to produce the twelve states   $^3{\bf 15}^{\prime P} _{1/2}$, $^3{\bf 15}^{\prime P} _{3/2}$, $^5{\bf 15} _{3/2}^P$, $^5{\bf 15} _{5/2}^P$,  $^3{\bf 15} _{1/2}^P$, $^3{\bf 15} _{3/2}^P$,  $^1{\bf 15} _{1/2}^P$,  $^3{\bf \bar 6} _{1/2}^P$, $^3{\bf \bar 6} _{3/2}^P$, $^3{\bf 3} _{1/2}^P$, $^3{\bf 3} _{3/2}^P$, and  $^1{\bf 3} _{1/2}^P$.  (The notation here is $^{2j_{\ell}+1}{\bf F}_J$, where $j_{\ell}$ is the light-quark spin, ${\bf F}$ is the flavor representation, and $J$ is the total spin of the state.) States with the same quantum numbers may mix; there are three mixing angles, for the two ${\bf 3}_{1/2}$ states, the two ${\bf 15}_{1/2}$ states, and the two ${\bf 15}_{3/2}$ states.   Heavy quark effective theory can be combined with the large $N_c$ formalism to produce an expansion in $1/N_c$ and $1/m_Q$ \cite{jenkins}.  The operators at order $1/(N_c m_Q)$ are

\begin{align}
O_7 & \equiv \frac{1}{N_cm_Q} S_c^i J_Q^i \nn \\
O_8 & \equiv \frac{1}{N_cm_Q} s^i J_Q^i \nn \\
O_9 & \equiv \frac{2}{N_c^2m_Q} t^a \{J_Q^i,G_c^{ia} \} \nn \\
O_{10} & \equiv \frac{1}{N_c^2m_Q} g^{ia} J_Q^i T_c^a
\end{align}
%%%%%%%%%%%%%%%%%%%%%%%%%%%%%
where $J_Q^i$ is the spin of the heavy antiquark. We find the mass relations

\begin{align}
^3{\bf \bar 6} _{3/2}^P - ^3{\bf \bar 6} _{1/2}^P & =  \frac{9}{10}(^5{\bf 15} _{5/2}^P - ^5{\bf 15} _{3/2}^P) - \frac{1}{2}(^3{\bf 15}^{\prime P} _{3/2} - ^3{\bf 15}^{\prime P} _{1/2}) + \mathcal{O}(1/N_c^2m_Q)  \\
^5{\bf 15}^P_{5/2}-^5{\bf 15}^P_{3/2} & = \frac{5}{3}\left(^3{\bf 15}^P_{3/2}-^3{\bf 15}^P_{1/2}\right) \\
^3{\bf \bar 6} _{3/2}^P - ^3{\bf \bar 6} _{1/2}^P & =  {\bf 6} _{3/2}^N - {\bf 6} _{1/2}^N
 + \mathcal{O}(1/N_c^2m_Q) \label{heavy}
\end{align}
%%%%%%%%%%%%%%%%%%%%%%%%%%%%%%%%%%%%%%%%
where ${\bf 6} _{1/2}^N$ and ${\bf 6} _{3/2}^N$ are the nonexotic heavy baryon  multiplets containing the $\Sigma_{c,b}$ and $\Sigma_{c,b}^*$, respectively.  In the charmed case, the mass splitting in eq. (\ref{heavy}) is 65.6 MeV \cite{pdg}.

In this paper, we have discussed a possible way of studying negative-parity exotic baryons using the $1/N_c$ expansion.  This formalism could be extended to investigate, for example, the decay widths of these states and $SU(3)$-breaking corrections to their masses.  As argued in \cite{stewart}, the ${\bf 3}_0^P$ multiplet, at least, may be stable against strong decays, so it is possible that someday these predictions may be tested experimentally.  

Shortly after the completion of this paper, Dan Pirjol and Carlos Schat posted \cite{pirjolschat}, which uses a very similar approach to pentaquarks, both light and heavy, in $1/N_c$.

\acknowledgments 
The author would like to thank Elizabeth Jenkins and Aneesh Manohar for helpful discussions, and Dan Pirjol for comments on a draft of this manuscript.
This work was supported in part by the
Department of Energy under Grant No.~DE-FG03-92-ER40701.

\newpage
\appendix

\section{Explicit Matrix Elements}

\begin{table}[h]
\begin{tabular}{|| c | c | c | c | c | c | c | c | c |}
  \hline
  &  $O_1$ &  $O_2$ & $O_3$ &  $O_4$ &  $O_5$ &  $O_6$ & $O_{11}$ \\ 
 & $N_c \openone$ & $\frac{1}{N_c}S_c^2$ &  $\frac{1}{N_c}s^iS_c^i$ & $\frac{1}{N_c}t^aT_c^a$ &  $\frac{1}{N_c^2}t^a \{ S_c^i, G_c^{ia} \}$ &  $\frac{1}{N_c^2}g^{ia}S_c^iT_c^a$ & $\frac{1}{N_c^3}S_c^2t^aT_c^a$ \\ \hline 
 ${\bf 3}_0^P$ & $N_c$ & $ \frac{3}{4N_c}$ & $- \frac{3}{4N_c}$ & $- \frac{N_c+6}{6N_c}$ & $- \frac{1}{2N_c^2}$ & $ \frac{N_c+6}{8N_c^2}$ & $-\frac{1}{8N_c^2}$ \\ \hline
  ${\bf 3}_1^P$ & $N_c$ & $ \frac{3}{4N_c}$ & $\frac{1}{4N_c}$ & $- \frac{N_c+6}{6N_c}$ & $- \frac{1}{2N_c^2}$ & $- \frac{N_c+6}{24N_c^2}$ & $-\frac{1}{8N_c^2}$ \\ \hline
 ${\bf \bar 6}^P_1$ & $N_c$ & $ \frac{3}{4N_c}$ & $ \frac{1}{4N_c}$ & $\frac{N_c-9}{12N_c}$ & $- \frac{3N_c+5}{4N_c^2}$ & $ \frac{N_c-9}{48N_c^2}$ &  $\frac{1}{16N_c^2}$ \\ \hline
${\bf 15}_0^P$ &  $N_c$ & $ \frac{3}{4N_c}$ & $- \frac{3}{4N_c}$ & $\frac{N_c+3}{12N_c}$ & $\frac{N_c+3}{4N_c^2}$ & $- \frac{N_c+3}{16N_c^2}$ & $\frac{1}{16N_c^2}$ \\ \hline
${\bf 15}_1^P$ & $N_c$ & $ \frac{7N_c+16}{4N_c^2}$ & $- \frac{7N_c+16}{12N_c^2}$ & $\frac{N_c^2-3N_c-25}{12N_c^2}$ & $- \frac{3N_c+23}{24N_c^2}$ & $\frac{-N_c+19}{48N_c^2}$ &  $\frac{7}{48N_c^2}$ \\ \hline
${\bf 15}_2^P$ &  $N_c$ & $ \frac{15}{4N_c}$ & $ \frac{3}{4N_c}$ & $\frac{N_c-15}{12N_c}$ & $- \frac{5(N_c+1)}{4N_c^2}$ & $ \frac{N_c-15}{16N_c^2}$ &  $\frac{5}{16N_c^2}$ \\ \hline
${\bf 15}^{\prime P} _1$ & $N_c$ & $ \frac{15}{4N_c}$ & $- \frac{5}{4N_c}$ & $\frac{N_c+9}{12N_c}$ & $\frac{3N_c+11}{4N_c^2}$ & $- \frac{5(N_c+9)}{48N_c^2}$ &  $\frac{5}{16N_c^2}$ \\ \hline
${\bf 10}_{3/2}^N$ & $N_c$ & $ \frac{2}{N_c}$ & $ \frac{1}{2N_c}$ & $\frac{N_c+5}{12N_c}$ & $\frac{3N_c+7}{6N_c^2}$ & $- \frac{N_c+5}{24N_c^2}$ &  $\frac{1}{6N_c^2}$ \\ \hline
${\bf 8}_{1/2}^N$ & $N_c$ & $ \frac{3(N_c-1)}{2N_c^2}$ & $- \frac{3(N_c-1)}{4N_c^2}$ & $\frac{N_c^2-10N_c+9}{12N_c^2}$ & $\frac{-3N_c^2+2N_c}{8N_c^3}$ & $ \frac{N_c^2+14N_c}{16N_c^3}$ & $\frac{1}{8N_c^2}$ \\ \hline
\end{tabular}
\caption{Matrix elements of singlet operators $O_1$ through $O_6$ and $O_{11}$ to order $1/N_c^2$.\protect \footnote{It should be noted that the {\it exact} matrix elements for the ${\bf 15}_1^P$ multiplet are in agreement with the results of \cite{pirjolschat}.  For Table I, $\frac{1}{N_c(N_c+1)}$ was expanded as $\frac{1}{N_c^2}-\frac{1}{N_c^3}+...$, and only terms up to order $1/N_c^2$ were kept.}} 
\end{table}

\begin{table}[h]
\begin{tabular}{|| c | c | c | c | c |}
  \hline
 & $O_7$ & $O_8$ & $O_9$ & $O_{10}$ \\ 
& $\frac{1}{N_cm_Q}S_c^iJ_Q^i$ &  $\frac{1}{N_cm_Q}s^iJ_Q^i$ & $\frac{2}{N_c^2m_Q}t^a \{ J_Q^i, G_c^{ia} \}$ & $\frac{1}{N_c^2m_Q}g^{ia}J_Q^iT_c^a$ \\ \hline
$^1{\bf 3}_{1/2}^P$ & 0 & 0 & 0 & 0  \\ \hline
$^3{\bf 3}_{3/2}^P$ & $\frac{1}{4N_cm_Q}$ & $\frac{1}{4N_cm_Q}$ & 0  & $- \frac{1}{24N_cm_Q}$ \\ \hline
$^3{\bf 3}_{1/2}^P$ & $- \frac{1}{2N_cm_Q}$ & $- \frac{1}{2N_cm_Q}$ & 0  & $\frac{1}{12N_cm_Q}$ \\ \hline
$^3{\bf \bar 6}_{3/2}^P$ & $\frac{1}{4N_cm_Q}$ & $\frac{1}{4N_cm_Q}$ & $- \frac{1}{4N_cm_Q}$  & $\frac{1}{48N_cm_Q}$ \\ \hline
$^3{\bf \bar 6}_{1/2}^P$ & $- \frac{1}{2N_cm_Q}$ & $- \frac{1}{2N_cm_Q}$ &  $\frac{1}{2N_cm_Q}$  & $- \frac{1}{24N_cm_Q}$ \\ \hline
$^1{\bf 15}_{1/2}^P$ & 0 & 0 & 0 & 0  \\ \hline
$^3{\bf 15}_{3/2}^P$ & $\frac{3}{8N_cm_Q}$ & $\frac{1}{8N_cm_Q}$ & $- \frac{1}{8N_cm_Q}$  & $\frac{1}{96N_cm_Q}$ \\ \hline
$^3{\bf 15}_{1/2}^P$ & $- \frac{3}{4N_cm_Q}$ & $-\frac{1}{4N_cm_Q}$ & $\frac{1}{4N_cm_Q}$  & $- \frac{1}{48N_cm_Q}$ \\ \hline
$^5{\bf 15}_{5/2}^P$ & $\frac{3}{4N_cm_Q}$ & $\frac{1}{4N_cm_Q}$ & $- \frac{1}{4N_cm_Q}$  & $\frac{1}{48N_cm_Q}$ \\ \hline
$^5{\bf 15}_{3/2}^P$ & $- \frac{9}{8N_cm_Q}$ & $- \frac{3}{8N_cm_Q}$ & $\frac{3}{8N_cm_Q}$  & $- \frac{1}{32N_cm_Q}$ \\ \hline
$^3{\bf 15}^{\prime P} _{3/2}$ & $\frac{5}{8N_cm_Q}$ & $- \frac{1}{8N_cm_Q}$ & $\frac{1}{8N_cm_Q}$  & $- \frac{1}{96N_cm_Q}$ \\ \hline
$^3{\bf 15}^{\prime P} _{1/2}$ & $- \frac{5}{4N_cm_Q}$ & $\frac{1}{4N_cm_Q}$ & $- \frac{1}{4N_cm_Q}$  & $\frac{1}{48N_cm_Q}$ \\ \hline
${\bf 6}_{3/2}^N$ & $\frac{1}{4N_cm_Q}$ & $\frac{1}{4N_cm_Q}$ & $- \frac{1}{4N_cm_Q}$  & $ \frac{1}{48N_cm_Q}$ \\ \hline
${\bf 6}_{1/2}^N$ & $- \frac{1}{2N_cm_Q}$ & $- \frac{1}{2N_cm_Q}$ &  $\frac{1}{2N_cm_Q}$  & $- \frac{1}{24N_cm_Q}$ \\ \hline
$^3{\bf 3}_{1/2}^P-^1{\bf 3}_{1/2}^P$ & $-\frac{\sqrt{3}}{4N_cm_Q}$ & $\frac{\sqrt{3}}{4N_cm_Q}$ & 0 & $-\frac{1}{2\sqrt{3}N_cm_Q}$ \\ \hline
$^3{\bf 15}_{1/2}^P-^1{\bf 15}_{1/2}^P$ & $\frac{1}{2\sqrt{2}N_cm_Q}$ &  $-\frac{1}{2\sqrt{2}N_cm_Q}$ & $-\frac{1}{2\sqrt{2}N_cm_Q}$  & $-\frac{1}{24\sqrt{2}N_cm_Q}$ \\ \hline
$^5{\bf 15}_{3/2}^P-^3{\bf 15}_{3/2}^P$ & $-\frac{\sqrt{5}}{8N_cm_Q}$ & $\frac{5}{8\sqrt{3}N_cm_Q}$ & $\frac{\sqrt{5}}{6\sqrt{5}N_cm_Q}$  & $\frac{\sqrt{5}}{96N_cm_Q}$  \\ \hline

\end{tabular}
\caption{Matrix elements of heavy operators $O_7$ through $O_{10}$ to order $1/N_cm_Q$.  The last three rows show off-diagonal matrix elements, which are related to the mixing angles.}
\end{table}

\newpage

\section{Flavor $SU(3)$ Multiplets}

\begin{figure}[h]
\PSbox{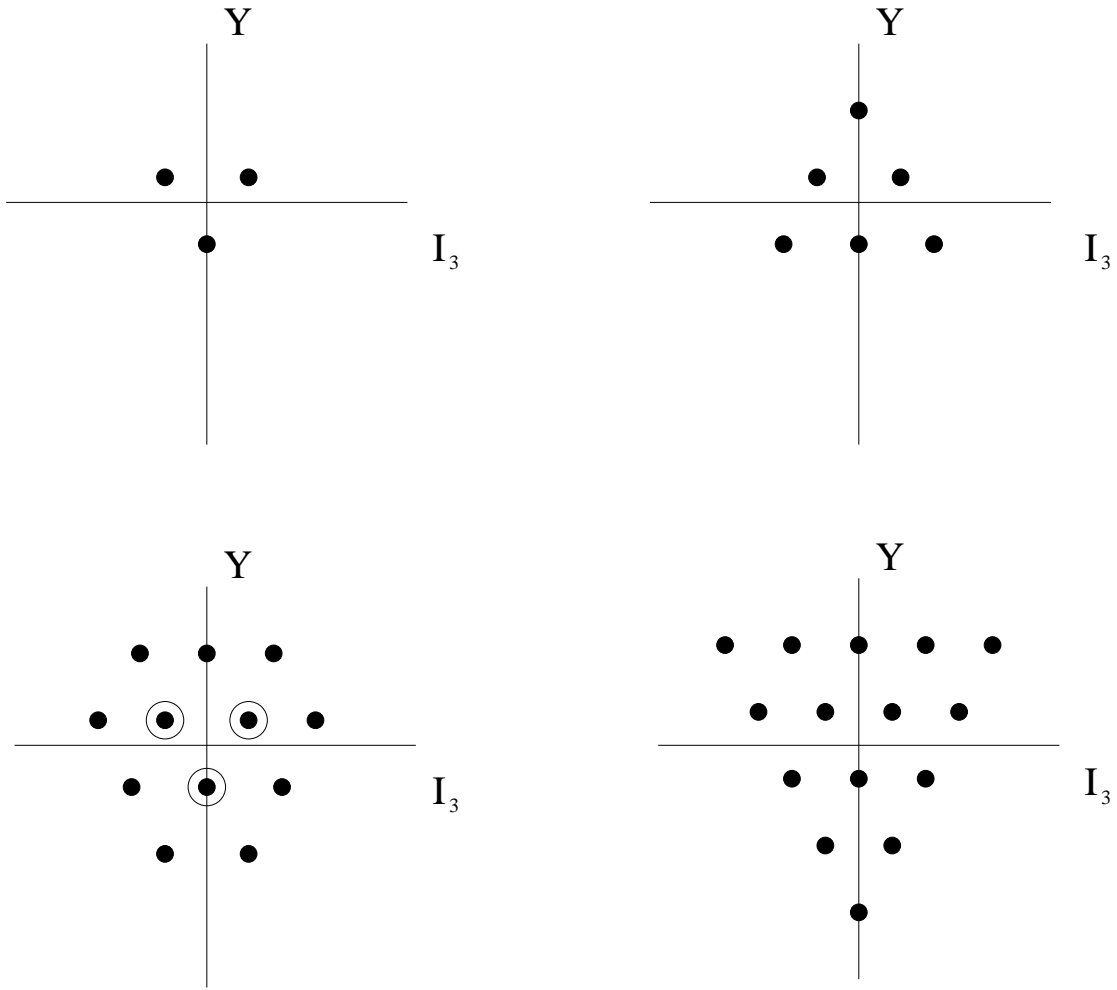 hoffset=10 voffset=30 hscale=70
vscale=70}{6.0in}{6.0in}
\caption{The four possible $SU(3)$ flavor multiplets for the negative-parity heavy pentaquarks.  The top row shows the ${\bf 3}^P$ and the ${\bf \bar 6}^P$; the bottom row shows the ${\bf 15}^P$ and the ${\bf 15}^{\prime P}$.}
\label{fig:c3}
\end{figure}

\end{document}